\documentclass[12pt]{article}

\usepackage{graphicx}

\begin{document}

\input{epsf.sty}

\begin{titlepage}

\begin{flushright}
IUHET-467\\
%hep-th/xxxxxxx
\end{flushright}
\vskip 2.5cm

\begin{center}
{\Large \bf Nonpolynomial Normal Modes of the Renormalization Group in the
Presence of a Constant Vector Potential Background}
\end{center}

\vspace{1ex}

\begin{center}
{\large B. Altschul\footnote{{\tt baltschu@indiana.edu}}}

\vspace{5mm}
{\sl Department of Physics} \\
{\sl Indiana University} \\
{\sl Bloomington, IN 47405 USA} \\

\end{center}

\vspace{2.5ex}

\medskip

\centerline {\bf Abstract}

\bigskip

We examine the renormalization group flow in the vicinity of the free-field
fixed point for effective field theories in the presence of
a constant, nondynamical vector potential background. The interaction with this
vector potential represents the simplest possible form of Lorentz violation.
We search for any
normal modes of the flow involving nonpolynomial interactions. For scalar
fields, the inclusion of the vector
potential modifies the known modes only through a change
in the field strength renormalization. For fermionic theories, where an infinite
number of particle species are required in order for nonpolynomial interactions
to be possible, we find no evidence for any analogous relevant modes. These
results are
consistent with the idea that the vector potential interaction, which may be
eliminated from the
action by a gauge transformation, should have no physical effects.

\bigskip 

\end{titlepage}

\newpage

In quantum theory, the electromagnetic field is incorporated through the use
of the vector potential $A^{\mu}$. While it is commonly held that only the
electric and magnetic fields are physically relevant, the vector potential
itself can contribute to the physics in curious ways~\cite{ref-aharonov}. If the
vector potential is in a pure gauge configuration, $A^{\mu}(x)=\partial^{\mu}
\Omega(x)$, then the electromagnetic field tensor $F^{\mu\nu}$ vanishes;
however, the nonzero $A^{\mu}$ still appears in the action, through the coupling
to charged matter. We shall examine the effects of this apparently trivial
coupling on the renormalization group (RG) flows for effective field theories.

The presence of a nonvanishing $A^{\mu}$ in the vacuum state is actually the
simplest possible form of Lorentz violation, and
the possibility of this and other forms of Lorentz violation has recently
received a significant amount of attention.
There has been a great deal of perturbative
work examining the general structure of the relevant quantum field
theories~\cite{ref-kost1,ref-kost2,ref-kost3} and the one-loop quantum
corrections~\cite{ref-jackiw1,ref-victoria,ref-chung1,ref-coleman,ref-chung2,
ref-kost4,ref-kost5}.
Small violations of Lorentz symmetry may arise in a low-energy effective field
theory as remnants of larger violations appearing in a fundamental theory at
higher energies. So it is natural to study Lorentz violation in the context of
an effective theory with a cutoff. We shall examine the renormalization group
flow for theories of this sort, in the vicinity of the Gaussian fixed
point. Our results will be nonperturbative in the Lorentz-violating parameter.
By studying the RG trajectories, we
shall be considering some further effects of quantum corrections.

In particular, we shall be concerned with any nonpolynomial interactions that
might arise in an effective field theory.
Halpern and Huang~\cite{ref-huang1,ref-huang2}, using the Wilsonian formulation
of the
RG~\cite{ref-wilson,ref-wegner,ref-hasenfratz}, have shown that in scalar field
theories,
there exist relevant nonpolynomial interactions; and Periwal~\cite{ref-periwal},
using exact RG methods~\cite{ref-polchinski}, has derived a
regularization-independent differential equation governing the
RG flow for these interactions.

This paper is organized as follows:
We first present a simplified and improved derivation of the differential
equation governing the linearized RG flow in the
scalar field case. Our treatment makes clear the role of tadpole loops in
determining the RG flow.
We then generalize this method to include the possibility of a constant,
nonvanishing vector potential, interacting with either bosons or fermions.
In the bosonic case, we find that the presence of the vector potential causes
the form of the nonpolynomial interactions to undergo an apparent change.
However, we also show that this change may be cancelled out by a redefinition of
the coupling constant. In the fermionic generalization, which requires the
presence of an infinite number of fermion species, we find no nonpolynomial
modes generated by the presence of the potential. We conclude with a discussion
of the significance of these results, as well as several subtle points related
to our calculations.

\subparagraph{Potential-free theory---}
We begin by re-examining the Halpern's and Huang's
original result. We shall enumerate all the
Feynman diagrams that contribute to the RG flow at lowest order and then derive
the partial differential equation governing the coupling constants' evolution.
We find that the dominant loop contributions come entirely from tadpole
diagrams. Although we use Feynman diagrams, our results are nonperturbative,
because we shall be summing an infinite number of diagrams, and the integration
momenta in the tadpole loops are correlated at all orders.

We first consider a theory with a single real scalar field; the generalization
to multiple scalar fields is simple. We shall investigate the
quantum corrections to this theory using bare perturbation theory.
The Euclidean action for the unrenormalized scalar theory is
\begin{equation}
S_{b}=\int d^{d}x\left[\frac{1}{2}(\partial\phi)^{2}+V_{b}\left(\phi^{2}\right)
\right],
\end{equation}
where the interaction Lagrange density $V_{b}$ is representable as a power
series in $\phi^{2}$. From this bare action, we may determine the renormalized
effective action, which arises from loop corrections.
We shall use a momentum cutoff $\Lambda$ to regulate the loop integrals
that arise in the calculation of the effective action.

\begin{figure}[t]
\epsfxsize=3in
\begin{center}
\leavevmode
\epsfbox{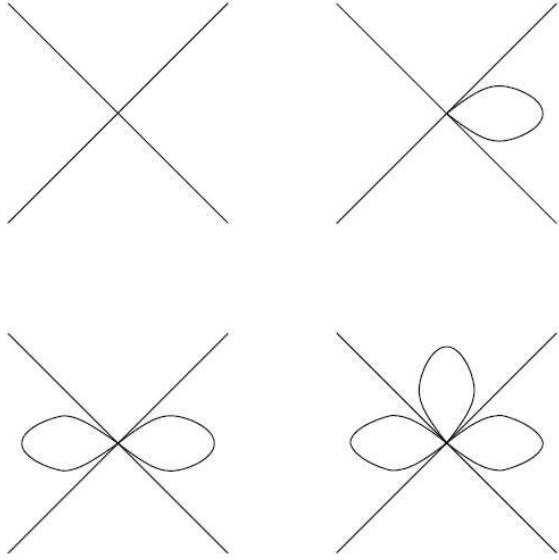}
\caption{Graphs contributing to the four-particle amplitude.\label{fig-graphs}}
\end{center}
\end{figure}

To determine the full effective action, we need to find the coefficient of each
effective vertex of the theory, and to determine the exact coefficient of an
effective $n$-particle vertex, we need to calculate the full $n$-particle
correlation function from the bare theory. However, we shall only be interested
in a restricted class of interactions---those which involve no derivatives (i.e.
are
independent of the external momenta) and which are at most linear in the bare
couplings. We may therefore make an approximation and omit some Feynman diagrams
in our determination of the $n$-particle correlation function.
When we neglect the momentum-dependent
interactions and the interactions that are nonlinear in $V_{b}$,
we find that the contributions to the effective $n$-particle amplitude
have a distinct form.  Written as a Feynman diagram, each of these
contributions contains a bare $(n+2k)$-point vertex and $k$ tadpole loops. The
lowest-order diagrams contributing to the four-particle amplitude are shown
in Fig.~\ref{fig-graphs}.

%\begin{figure}[t]
%\begin{center}
%\includegraphics[height=0.4\textwidth,angle=0]{figure-loops.pdf}
%\caption{Graphs contributing to the four-particle amplitude.\label{fig-graphs}}
%\end{center}
%\end{figure}

We can sum up all diagrams of this sort without difficulty~\cite{ref-halpern}.
Each loop
contributes a factor of $\frac{1}{2}D_{F}(0)$, where $D_{F}(x-y)$ is the
Feynman propagator for a massless scalar particle; $D_{F}$ differs from minus
the
inverse Laplacian only through its large-momentum regulation. The factor of 2
is the symmetry factor for the loop. There is also an additional symmetry factor
of $k!$ corresponding to interchanging of the $k$ loops.

We want to express the effective coupling constants in a dimensionless fashion,
so we write the effective interaction action as
\begin{equation}
S_{b,int}=\int d^{d}x\,\Lambda^{d}U_{b}\left[\Lambda^{-(d-2)/2}
\phi\right].
\end{equation}
The function $U_{b}\left[\Lambda^{-(d-2)/2}\phi\right]$ also depends upon
$\Lambda$
as a parameter; this parametric $\Lambda$-dependence represents the
dimensionally anomalous behavior of the effective potential.
When we total up the
combinatorial factors, we find that, with the approximations mentioned above,
the effective potential is given by
\begin{equation}
\label{eq-diagsumb}
\Lambda^{d}U_{b}\left[\Lambda^{-(d-2)/2}\phi
\right]=\sum_{k=0}^{\infty}\frac{1}{k!}\left[\frac{1}{2}D_{F}(0)\frac{\partial
^{2}}{\partial\phi^{2}}\right]^{k}V_{b}\left(\phi^{2}\right).
\end{equation}
In dimensions $d>2$, the regulated value of $D_{F}(0)$ has the form $C_{b}
\Lambda^{d-2}$, for some constant $C_{b}$. The explicit value of $C_{b}$
is given in~\cite{ref-huang1}. (When $d=2$, the dependence on $\Lambda$ is
logarithmic, but analogous results continue to hold.)
If we insert this expression into
(\ref{eq-diagsumb}), we have
\begin{equation}
\label{eq-diagexpb}
\Lambda^{d}U_{b}\left[\Lambda^{-(d-2)/2}\phi
\right]=\exp\left[\frac{1}{2}C_{b}\Lambda^{d-2}\frac{\partial
^{2}}{\partial\phi^{2}}\right]V_{b}\left(\phi^{2}\right).
\end{equation}
If we apply the operator $\Lambda\frac{\partial}{\partial\Lambda}$ to both sides
of (\ref{eq-diagexpb}) and then use (\ref{eq-diagexpb}) to rearrange the
right-hand side, we find
\begin{equation}
\label{eq-periwalDE}
\Lambda\frac{\partial U_{b}}{\partial\Lambda}+dU_{b}-\frac{d-2}{2}
\Lambda^{-(d-2)/2}\phi U_{b}'\left[\Lambda^{-(d-2)/2}\phi\right]=\frac{d-2}
{2}C_{b}U_{b}''\left[\Lambda^{-(d-2)/2}\phi\right],
\end{equation}
which is essentially just Periwal's differential equation. We can see that the
right-hand side, which determines the quantum corrections to the RG flow, has
come entirely from tadpole contributions.
The solutions of the differential equation (\ref{eq-periwalDE}) with power-law
dependences on $\Lambda$---that is, with $\Lambda\frac{\partial U_{b}}{\partial
\Lambda}=-\lambda U_{b}$--- are eigenmodes of the RG flow near the Gaussian
fixed point.
When $\lambda>0$, these potentials describe theories that are asymptotically
free, with much stronger scale dependences than are seen in superficially
renormalizable
theories.
All these relevant directions correspond to nonpolynomial
interactions.

The complex perturbative structure of these
theories is outlined in~\cite{ref-periwal}. Of particular importance is the fact
that the anomalous
dimension $\lambda$ describes only how the
potential scales with respect to the cutoff $\Lambda$; it does not
describe the scaling of any correlation functions with respect to their
external momenta.

\subparagraph{Bosonic theory with $A^{\mu}$---}
We now generalize the theory to include a constant, nondynamical electromagnetic
potential $A^{\mu}(x)=a^{\mu}$, interacting with a complex scalar field $\Phi$.
The vector $a$ represents a preferred direction in spacetime and is the source
of the breaking of (Euclidean) Lorentz invariance.
The action becomes
\begin{equation}
S_{b}=\int d^{d}x\left\{\left[\left(\partial_{\mu}-ia_{\mu}\right)\Phi^{*}
\right]\left[\left(\partial^{\mu}+ia^{\mu}\right)\Phi\right]
+V_{b}\left(2|\Phi|^{2}\right)\right\}.
\end{equation}
The modified Lagrange density involves two new terms. These are the interaction
term $ia^{\mu}\left[\left(\partial_{\mu}\Phi^{*}\right)\Phi-
\Phi^{*}\left(\partial_{\mu}\Phi\right)\right]$
and a shift in the mass term $a^{2}|\Phi|^{2}$.

The preferred vector $a$ can be absorbed into the definition of the field,
according to
\begin{equation}
\label{eq-redef}
\Phi\rightarrow e^{-ia\cdot x}\Phi.
\end{equation}
If the theory is defined perturbatively in $a$, then this gauge transformation
removes
$a$ from the theory entirely~\cite{ref-kost2}. (However, in the presence of
multiple particle species, interacting with multiple vector potentials,
the differences between the values of $a$ corresponding to different
species can be observable~\cite{ref-kost6,ref-kost7,ref-kost8}.)
Since $a$ can be removed in this fashion, the operator
$ia^{\mu}\left[\left(\partial_{\mu}\Phi^{*}\right)\Phi-
\Phi^{*}\left(\partial_{\mu}\Phi\right)\right]$
cannot receive quantum corrections at any order in perturbation theory
and therefore cannot
have a nonzero anomalous dimension. (This is explicitly verified to one-loop
order in~\cite{ref-kost4} for the
analogous fermionic theory that we shall examine later in this paper.)
The RG flow of $a$ is
thus purely classical. We may therefore write $a$ as $a=\hat{a}\theta
\Lambda$,
where $\theta$ is a scale-independent number, and $\hat{a}$ is a unit vector in
the direction of $a$.

Although $a$ can be eliminated by a field redefinition, there are still some
subtleties associated with it. If the magnitude of $a$ is too large, then the
quantization of the theory encounters difficulties~\cite{ref-kost3}, unless the
field redefinition is performed before the quantization procedure. Since we
are not eliminating $a$ before we quantize the theory, we must insist that the
magnitude of $a$ be small compared to some relevant mass scale, which we shall
take to be the cutoff $\Lambda$. This implies that $\theta\ll 1$.

The field redefinition (\ref{eq-redef}) does not leave the renormalization
prescription invariant (a consequence of the familiar fact
that a momentum cutoff regulator is not gauge invariant),
so it is conceivable that the Halpern-Huang modes may
be modified by the presence of this interaction. 
The key quantity that
determines the structure of these modes is $D_{F}(0)$. In four dimensions, and
in the presence of $a$, this becomes
\begin{equation}
D_{F}(0)=\int_{|p|<\Lambda}\frac{d^{4}p}{(2\pi)^{4}}\frac{1}{(p-a)^{2}}.
\end{equation}
This modification of $D_{F}(0)$ is the only effect of $a$ that we shall
consider.

The modified $D_{F}(0)$ may be evaluated by separating the integration variable
$p$ into portions
$p_{\parallel}$ and $p_{\perp}$, which are parallel and normal to $a$,
respectively. Then the propagator at zero separation becomes
\begin{eqnarray}
D_{F}(0) & = & \frac{1}{(2\pi)^{4}}\int_{-\Lambda}^{\Lambda}dp_{\parallel}
\int_{0}^{\sqrt{\Lambda^{2}-p_{\parallel}^{2}}}dp_{\perp}\frac{4\pi p_{\perp}
^{2}}{p_{\perp}^{2}+(p_{\parallel}-a)^{2}} \\
& = & \frac{\Lambda^{2}}{4\pi^{3}}\left[\frac{\pi}{2}-\int_{-1}^{1}dy\,
(y-\theta)\left(\tan^{-1}\frac{\sqrt{1-y^{2}}}{y-\theta}\right)\right],
\end{eqnarray}
where $y=p_{\parallel}/\Lambda$.
For $\theta=0$, $D_{F}(0)=\Lambda^{2}/16\pi^{2}$; for nonvanishing $a$, the
integral cannot be evaluated exactly.

Although we cannot obtain a closed-form expression for $D_{F}(0)$ when $\theta$
is nonvanishing, we can define a new constant
\begin{equation}
C_{b}(\theta)=\frac{D_{F}(0)}
{\Lambda^{2}}=\frac{1}{16\pi^{2}}+\frac{3}{32\pi^{2}}\theta^{2}+{\cal O}(\theta
^{4}).
\end{equation}
Then we may repeat our earlier arguments and obtain the linearized RG equations
in the presence of $a$. The only essential difference (aside from those
differences following from the presence
of two real scalar fields) involves the replacement $C_{b}\rightarrow C_{b}
(\theta)$. So
the normal modes of the RG flow in the vicinity of the free field fixed point
are exactly as found by Halpern and Huang, except that they involve the modified
$C_{b}(\theta)$.

The nonpolynomial normal modes are therefore described by
\begin{equation}
V_{b}\left(2|\Phi|^{2}\right)=\Lambda^{d}U_{b}\left[\sqrt{2}\Lambda^{-(d-2)/2}
\Phi\right],
\end{equation}
where $U_{b}$ is the nondimensionalized potential given by
\begin{equation}
U_{b}(y)=gM\left[\frac{\lambda-d}{d-2};1;\frac{y^{2}}{2C_{b}(\theta)}\right];
\end{equation}
here, $g$ is a coupling constant, and $M(\alpha;\beta;z)$ is the confluent
hypergeometric (Kummer) function~\cite{ref-abramowitz}
\begin{equation}
M(\alpha;\beta;z)=1+\frac{\alpha}{\beta}\frac{z}{1!}+\frac{\alpha(\alpha+1)}
{\beta(\beta+1)}\frac{z^{2}}{2!}+\cdots.
\end{equation}

We see that, in the presence of the background potential $a$, the functional
forms of the normal modes in fact differ from the forms they take when $a$
vanishes.
However, this difference is not physically meaningful. It is clear that the
difference appears only in the field strength renormalization. By changing the
normalization of the fields, we may absorb the ratio $C_{b}(0)/C_{b}(\theta)$
into the
coupling constant $g$. Alternatively, we might change the renormalization
prescription slightly, cutting off the $p$-integration in $D_{F}(0)$ at
$|p|=\Lambda\sqrt{C_{b}(0)/C_{b}(\theta)}$, rather than simply at $|p|=\Lambda$.
Such a
change in the cutoff will not affect the theory, except through a change in the
coupling. This is all in keeping with the results found in~\cite{ref-halpern},
where
the field strength renormalization in the presence of nonpolynomial potentials
is studied in detail.

That the change in the form of the potential can be eliminated by rescaling the
couplings is unsurprising, in light of the perturbative triviality of
the interaction we have introduced. However, it is reassuring to see that the
expected triviality persists in these nonperturbative results. Moreover, we can
further illuminate the common structure of this triviality by looking at the
behavior of the coupling constant $g$ in another particular context. As we have
seen, the field strength renormalization that restores the potential to the
Halpern-Huang form involves a finite multiplicative shift in the coupling. This
is similar to what we see when we use the gauge transformation (\ref{eq-redef})
to remove the operator
$ia^{\mu}\left[\left(\partial_{\mu}\Phi^{*}\right)\Phi-
\Phi^{*}\left(\partial_{\mu}\Phi\right)\right]$
from an otherwise free Lagrange density. As well as eliminating the interaction
of $a$ with the charged current, (\ref{eq-redef}) also introduces a shift in the
mass parameter of the theory. So in this case also, the elimination of the
nonstandard
interaction changes the coupling constant of the theory by a finite factor.

\subparagraph{Fermionic theory---}
We shall now turn our attention to the analogous fermionic theories. In order to
generalize the bosonic case to include a nondynamical vector potential, we
needed to work with a complex scalar field or, equivalently, two real scalar
fields. However, our fermionic generalization necessarily requires that we
introduce an {\em infinite} number of fields. Otherwise, the anticommutivity of
the fermion fields will prevent the existence of nonpolynomial potentials. So
we see that a generalization to fermion systems will necessarily be rather
complicated.

The presence of a nonvanishing $a$ (or some other preferred vector) is also
necessary for any generalization to fermions.
The curious RG flows in the Halpern-Huang interaction directions arise from the
factors of $D_{F}(0)=C_{b}\Lambda^{d-2}\neq0$
that appear in loop diagrams. In a Lorentz-invariant theory, the corresponding
quantity for Dirac fermions,
\begin{equation}
S_{F}(0)=i\int\frac{d^{d}p}{(2\pi)^{d}}\frac{\not\! p}{p^{2}},
\end{equation}
vanishes. However, in a theory containing one or more
preferred vectors, $S_{F}(0)$ need not be zero (although its trace
still must vanish). Then there may again be nontrivial relevant directions near
the Gaussian fixed point.

So we shall consider a fermionic theory with unrenormalized Euclidean
action
\begin{equation}
S_{f}=\int d^{d}x\left[\psi^{\dag}\left(\!\not\!\partial\,+i\!\!\not\!a\,\right)
\psi+V_{f}\left(i\psi^{\dag}\psi\right)\right].
\end{equation}
$\psi$ and $\psi^{\dag}$ are Grassmann-valued fields, and
the $D$-dimensional Euclidean Dirac matrices are anti-hermitian and obey
$\{\gamma_{\alpha},\gamma_
{\beta}\}=-2\delta_{\alpha\beta}I_{D}$, where $I_{D}$ is the
identity matrix in spinor space. (We must eventually take the
limit $D\rightarrow\infty$, to represent the infinite number of particle
species.) Complex conjugation does not reverse the
order of the Grassmann numbers---$(\alpha\beta)^{*}=\alpha^{*}\beta^{*}$.
The function $V_{f}$ is real
valued and must be expandable as a power series. 
With these conventions (which follow~\cite{ref-ramond}), the action is real:
$S_{f}^{*}=S_{f}$.
We shall again regulate our theory with a momentum cutoff $\Lambda$ for any loop
integrals.
All the arguments regarding the perturbative triviality of $a$ and its effects
on the quantization of the theory
apply exactly as in the bosonic case.

We may now directly generalize the methods we used to arrive at the differential
equation governing the RG flow for bosons.
We again evaluate the relevant
propagator at zero separation. In the presence of $a$, we have, for $d>1$,
\begin{equation}
\label{eq-SF}
S_{F}(0)=i\int_{|p|<\Lambda}\frac{d^{d}p}{(2\pi)^{d}}\frac{\not\! p\,-\not\! a}
{(p-a)^{2}}=i\!\not\!\hat{a}\,C_{f}(\theta)\Lambda^{d-1}.
\end{equation}
In four dimensions, the constant $C_{f}(\theta)$ is given by
\begin{eqnarray}
C_{f}(\theta) & = &
\frac{1}{4\pi^{3}}\left[-\frac{\pi}{2}\theta-\int_{-1}^{1}dy\,
(y-\theta)^{2}\left(\tan^{-1}\frac{\sqrt{1-y^{2}}}{y-\theta}\right)\right] \\
& = & -\frac{1}{32\pi^{2}}\theta+{\cal O}(\theta^{3}).
\end{eqnarray}
Again, this modification of the zero-separation propagator is the only effect
of $a$ that we shall consider.

The fact that $S_{F}(0)$ is not zero might allow for the existence of novel,
asymptotically free theories analogous to those found in the bosonic case.
Following this analogy, we look for an effective action with an interaction term
of the form
\begin{equation}
S_{f,int}=\int d^{d}x\,\Lambda^{d}U_{f}\left[\Lambda^{-(d-1)}i\psi^{\dag}\psi
\right].
\end{equation}
Again, by assuming that $S_{f,int}$ has this form, we are neglecting
momentum-dependent interactions. Moreover, we are also neglecting any
interactions that have a nontrivial matrix structure in spinor space, such as
$(\psi^{\dag}\gamma\psi)^{2}$. Finally, we shall again neglect terms that are
nonlinear in the bare couplings, since these do not affect the structure near
the free-field fixed point.

We must now enumerate the contributing loop diagrams. Because the fermions have
a nontrivial spinor structure, this is more complicated than in the scalar field
case. However, we may utilize that fact that $D\rightarrow\infty$ to simplify
our calculation.
A nonvanishing fermion loop is a contraction of $\left(\psi^{\dag}\psi
\right)^{2m}$, such that the contractions form a single closed cycle. (We shall
refer to this as a loop of size $2m$.)
Each such loop contributes a factor of $D$, arising from a trace over the spinor
space. As $D\rightarrow\infty$, only diagrams with loops of this form need be
considered; all other topologies give contributions small in comparison.
(However, those other topologies are critically important for fixed, finite $D$.
In such a situation, the terms we have neglected ensure that there are
no contributions from terms in $V_{f}$ with more than $D$ fermion
operators.) There
are $(2m-1)!$ ways to perform the contraction in constructing a loop of size
$2m$; this is equivalent to a symmetry
factor of $2m$, corresponding to the $2m$ cyclic permutations of the
$\psi^{\dag}\psi$ pairs that leave the contraction invariant. The value of a
loop of size $2m$ is therefore $-\frac{1}{2m}{\rm tr}\left[S_{F}(0)\right]
^{2m}.$

When we assemble the diagrams of interest, the loops of size $2m$ exponentiate
as in the
bosonic case, independently the loops of other sizes. We find that the effective
interaction is therefore
\begin{equation}
\Lambda^{d}U_{f}\left[\Lambda^{-(d-1)}i\psi^{\dag}\psi\right]=\exp\left\{-D
\sum_{m=0}^{\infty}(-1)^{m}\frac{1}{2m}\left[C_{f}(\theta)\right]^{2m}\Lambda
^{2m(d-1)}\frac{
\partial^{2m}}{\partial\left(\psi^{\dag}\psi\right)^{2m}}\right\}V_{f}\left(i
\psi^{\dag}\psi\right).
\end{equation}
This leads to a differential equations analogous to (\ref{eq-periwalDE}); in
terms of $x=\Lambda^{d-1}i\psi^{\dag}\psi$, it is
\begin{equation}
\label{eq-fermiDE}
\Lambda\frac{\partial U_{f}}{\partial\Lambda}+dU_{f}-(d-1)xU_{f}'(x)=-D(d-1)
\left\{\sum_{m=0}^{\infty}\left[C_{f}(\theta)\right]^{2m}\frac{\partial^{2}}
{\partial x^{2}}\right\}U_{f}(x).
\end{equation}
If we take $U_{f}$ to be an eigenfunction of the RG flow, with $\Lambda\frac
{\partial U_{f}}{\partial\Lambda}=-\lambda U_{f}$, and act on both sides of the
differential equation with $\left[1-C_{f}(\theta)^{2}\frac
{\partial^{2}}{\partial x^{2}}\right]$, then we have
\begin{equation}
\left\{\left[1-C_{f}(\theta)^{2}\frac{\partial^{2}}{\partial x^{2}}\right]\left[
(\lambda-d)+(d-1)x\frac{\partial}{\partial x}\right]-(d-1)D\right\}U_{f}(x)=0.
\end{equation}

This third-order differential equation has two solutions that are representable
as power series about $x=0$. They can be expressed as generalized hypergeometric
functions~\cite{ref-gradsteyn}
\begin{eqnarray}
f_{1}^{\lambda}(x) & = & _{1}F_{2}\left[\frac{1}{2}\left(\frac{\lambda-d}{d-1}-
D\right);
\frac{1}{2}\left(\frac{\lambda-d}{d-1}+1\right),\frac{1}{2};\frac{x^{2}}{4
C_{f}(\theta)^{2}}\right] \\
f_{2}^{\lambda}(x) & = & x\left\{_{1}F_{2}\left[\frac{1}{2}\left(\frac{\lambda-
d}{d-1}-D+1\right);
\frac{1}{2}\left(\frac{\lambda-d}{d-1}+2\right),\frac{3}{2};\frac{x^{2}}{4
C_{f}(\theta)^{2}}\right]\right\}.
\end{eqnarray}
The function $_{1}F_{2}(\alpha;\beta,\gamma;z)$ (not to be confused with the
more frequently occurring $_{2}F_{1}$) is defined to be
\begin{equation}
_{1}F_{2}(\alpha;\beta,\gamma;z)=\sum_{i=0}^{\infty}\frac{\alpha(\alpha+1)\cdots
(\alpha+i-1)}{\beta(\beta+1)\cdots(\beta+i-1)\cdot\gamma(\gamma+1)\cdots
(\gamma+i-1)}\frac{z^{i}}{i!}.
\end{equation}

The functions $f_{1}^{\lambda}(x)$ and $f_{2}^{\lambda}(x)$ do not have
well-defined limits as $D\rightarrow\infty$ for any finite $\lambda$. Therefore,
there can be no nonpolynomial interactions with power-law coupling constant
flows (either relevant or irrelevant) generated in this manner. This is again
in keeping with the triviality of $a$.

\subparagraph{Discussion---}
There are still some subtle points to be addressed regarding our discussion of
the fer\-mi\-on\-ic case.
We have neglected a large number of diagrams that could generate
other interesting effects, and we must keep in mind that our
differential equation (\ref{eq-fermiDE}) is valid only in the limit of infinite
$D$. In fact, as $D\rightarrow\infty$, the equation itself does not have a
well-defined
limit. However, the solutions may have meaningful limits, if
$\lambda\rightarrow\infty$ also; this would appear to
correspond to an infinitely rapid coupling constant flow. Specifically, if
$\frac{\lambda-d}{d-1}=cD$ as $D$ approaches infinity, then
$f_{1}^{\lambda}(x)$ and $f_{2}^{\lambda}(x)$ approach
\begin{eqnarray}
\label{eq-f1infty}
f_{1}^{\lambda}(x) & \rightarrow & \cosh\left(\sqrt{\frac{c-1}{c}}\frac{x}{C_{f}
(\theta)}\right) \\
\label{eq-f2infty}
f_{2}^{\lambda}(x) & \rightarrow & C_{f}(\theta)\sqrt{\frac{c}{c-1}}\sinh\left(
\sqrt{\frac{c-1}{c}}\frac{x}{C_{f}(\theta)}\right).
\end{eqnarray}

Since (\ref{eq-f1infty}) and (\ref{eq-f2infty}) apparently correspond to
interactions with
infinitely rapid coupling constant flow, we might be inclined to conclude that
no interaction $U_{f}(x)=g\exp(\alpha x)$ may be generated by radiative
corrections, for the following reason: If such an interaction existed at some
scale, then a finite shift in the cutoff $\Lambda$ could generate a new
interaction that was infinitely strong, and hence unphysical. One might then
further conclude that we had found a nontrivial effect caused by the presence
of $a$. However, both conclusions are erroneous. The second conclusion is
incorrect, because we have no {\em a priori} reason to believe that the
exponential interactions in question are allowed even in the absence of a
nonvanishing $a$; there might be other effects that prohibit their generation,
unrelated to those discussed here. Moreover, even the first conclusion, that the
potentials are forbidden, is incorrect. We have only considered the
linearized coupling constant flows in the vicinity of the free-field fixed
point. The RG flow for the exponential potential $U_{f}(x)=g\exp(\alpha x)$
necessarily carries the theory into the region of strong
coupling, far from the fixed point, and so our treatment does not apply. We
therefore cannot conclude anything about whether or not these interactions are
allowed, and the question of whether the presence of $a$ may generate
interactions of this particular form remains open.

Thus far, we have worked entirely in Euclidean space. If we transform these
theories into Minkowski spacetime, we can draw some additional conclusions about
the physics. An
analytic continuation of $C_{f}(\theta)^{2}$ to Minkowski spacetime gives a
result proportional to
${\rm sgn}(a^{2})$, so, for lightlike $a$, there is clearly no possibility for
any nonpolynomial interaction to be generated by the means we have discussed.
This provides a partial answer to the question raised at the end of the previous
paragraph.

Our entire discussion has been based on the use of a regulator that is not gauge
invariant. This is not problematic in this instance, since the gauge field
$A^{\mu}$ is constant and nondynamical; but it is questionable whether our
methods would still be applicable in the presence of a nontrivial, quantized
electromagnetic
field. If we were required to use a gauge-invariant regulator, then $a$ would
certainly
not generate a change in the RG flow for any theory. However, since we have
identified no nontrivial changes to either the bosonic or fermionic theories,
our results are entirely consistent with this form of gauge invariance.

In this paper, we have provided some further confirmation of the triviality of
$a$---the Lorentz-violating parameter corresponding to the vacuum expectation
value of the vector potential $A^{\mu}$. We have performed nonperturbative RG
calculations in the presence of $a$, finding results entirely consistent with
the triviality of this term. In the bosonic case, the only changes to the
nonpolynomial Halpern-Huang
interactions may be eliminated by a rescaling of the couplings.
For the analogous fermionic system, there are no nonpolynomial modes in the
vicinity of the free-field fixed point that possess power-law coupling constant
flows. These results are consistent with the idea that the gauge parameter $a$
should have no measurable effects on any theory.

\section*{Acknowledgments}
The author is grateful to K. Huang, V. A. Kosteleck\'{y}, and T. Miller for
their helpful discussions.
This work is supported in part by funds provided by the U. S.
Department of Energy (D.O.E.) under cooperative research agreement
DE-FG02-91ER40661.

\end{document}